\documentclass[aps,twocolumn,superscriptaddress,showpacs,dvips]{revtex4}
\usepackage{feynmp}
\usepackage{amssymb}
\usepackage{epsfig}
\usepackage{graphicx}
\usepackage{subfigure}
\usepackage{hyperref}
\usepackage{bbm}

\begin{document}

\title{Nature of the antiferromagnetic quantum phase transition on the honeycomb lattice}

\author{Jing-Rong Wang}
\affiliation{Department of Modern Physics, University of Science and
Technology of China, Hefei, Anhui 230026, P. R. China}
\author{Guo-Zhu Liu}
\affiliation{Max Planck Institut f$\ddot{u}$r Physik komplexer
Systeme, D-01187 Dresden, Germany} \affiliation{Department of Modern
Physics, University of Science and Technology of China, Hefei, Anhui
230026, P. R. China}
\author{Stefan Kirchner}
\affiliation{Max Planck Institut f$\ddot{u}$r Physik komplexer
Systeme, D-01187 Dresden, Germany} \affiliation{Max Planck Institut
f$\ddot{u}$r chemische Physik fester Stoffe, D-01187 Dresden,
Germany}

\begin{abstract}
We address the nature of the antiferromagnetic quantum phase
transition that separates a semimetal from an antiferromagnet in the
repulsive Hubbard model defined on the honeycomb lattice. At the
critical point, the fermions acquire an anomalous dimension $\eta$
due to their strong coupling to the fluctuations of the order
parameter $\phi$. The finite $\eta$ in turn induces a singular
$\phi^{4}$ term and a non-analytical spin susceptibility signaling
the breakdown of Hertz's $\phi^{4}$ theory. As a result, the
continuous antiferromagnetic quantum phase transition is internally
unstable and turns into a first order transition.
\end{abstract}


\maketitle

%%%%%%%%%%%%%%%%%%%%%%%%%%%%%Main Body%%%%%%%%%%%%%%%%%%%%%%%%%%%%%%%%%%%%%

Quantum phase transitions are believed to be present in a wide
variety of correlated quantum matter ranging from strongly
correlated electron systems to ultracold quantum gases on engineered
lattices \cite{understand}. Such quantum phase transitions occur at
zero temperature as a function of a non-thermal tuning parameter
\cite{Sachdevbook, understand, Vojta, Sachdev}. Understanding the
nature of possible quantum phase transitions of a given system and
describing the ensuing physical behaviors present a great challenge
of modern materials science. In particular, the possible
zero-temperature phase transitions in various models of interacting
electrons on the two-dimensional honeycomb lattice have recently
generated intense theoretical and numerical interests
\cite{Herbut06, Herbut09, Honerkamp, Sorella, Hermele, Wangfa, Ran,
Wen, Meng, Sachdev, Fritz}. While it is well-established that the
ground state of the repulsive Hubbard model is a semimetal (SM) at
weak coupling and a Mott-insulating antiferromagnet (AFM) at strong
coupling, the nature of the transition between these two phases is
still not well understood.

There are generally two scenarios for the transition from a SM to an AFM.
Herbut \cite{Herbut06, Herbut09} studied the repulsive Hubbard model
on a honeycomb lattice by means of a renormalization group (RG)
method in the large-$N$ limit and argued that the strong on-site
interaction drives a direct, continuous SM-AFM quantum phase
transition, which is in qualitative agreement with the results
obtained by a functional RG calculation \cite{Honerkamp} and early
quantum Monte Carlo (MC) simulations \cite{Sorella}. Treatments of
the Hubbard models based on auxiliary-particle methods on the
contrary seem to indicate that various exotic quantum spin liquid
phases may exist between the SM and AFM phases \cite{Hermele, Wangfa,
Ran, Wen}. Particularly, a recent quantum MC simulation \cite{Meng}
found an intermediate gapped spin liquid phase. These two
scenarios are apparently contradictory, and no consensus has been
reached so far.

%Since there is little doubt regarding the ground state
%properties at weak and strong coupling limits, here we focus on the
%intermediate coupling region and perform a detailed analysis of the
%SM-AFM quantum critical point.

The theoretical study of quantum phase transitions was pioneered by
Hertz \cite{Hertz}, who developed a field-theoretic approach to
describe the (anti-)ferromagnetic phase transition in an itinerant
electron system. In this approach, the fermionic degrees of freedom
are completely integrated out, yielding an effective $\phi^4$ action
soley in terms of an order parameter $\phi$. Hertz's theory and its
generalizations \cite{Millis, Moriya} have been applied to numerous
quantum phase transitions \cite{Sachdevbook, Vojta}. Recently,
however, the general validity of this approach has been called into
question \cite{Vojta, Belitz, Abanov, Chubukov} as it has become
clear that it is not always eligible to integrate out all fermionic
degrees of freedom in an itinerant electron system. Indeed, the
singular interaction between gapless itinerant fermions and order
parameter fluctuations can lead to important features that are
beyond a simple $\phi^4$ description \cite{Vojta, Belitz, Abanov,
Chubukov}.

In this Letter we address the nature of the quantum phase transition
between an antiferromagnet and a semimetal on the honeycomb lattice.
We approach the problem by assuming that the SM-AFM phase transition
is a continuous one and then examining whether it is stable or not.
To this end, we maintain the fermionic degrees of freedom of the
system, described as massless Dirac fermions, within a Yukawa-type
field theory to describe the SM-AFM quantum critical point. This
allows for a careful study of the singular interaction between Dirac
fermions and order parameter fluctuations.

We first compute the fermion anomalous dimension after incorporating
the damping effects of both Dirac fermions and order parameter. We
then calculate the coefficient of an effective quartic $\phi^{4}$
term and the spin susceptibility, and show that both of these two
quantities are non-analytical. As it turns out, the non-analytical
spin susceptibility leads to an effective energy of $\phi$ that has
a minimum at a finite momentum $q_c$, so a nonzero AFM order
parameter is generated at the transition point. This order parameter
is nonuniform and spatially modulated as $\langle \phi \rangle =
\phi_0 e^{\mathbf{q}_c \cdot \mathbf{x}}$.  All these unusual
behaviors indicate that, the original Hertz $\phi^{4}$ theory breaks
down and the Dirac fermions and order parameter fluctuations should
be treated equally. As we demonstrate below, the putative continuous
SM-AFM quantum phase transition is destroyed by the critical order
parameter fluctuations and consequently becomes first order.

We begin with the half-filled Hubbard model
\begin{equation}
\mathcal{H} = -\sum_{<i,j>}t c_{i\sigma}^{\dagger}c_{j\sigma} +
\sum_{i}U\left(n_{i\uparrow}-\frac{1}{2}\right)
\left(n_{i\uparrow}-\frac{1}{2}\right)
\end{equation}
defined on a two-dimensional honeycomb lattice. Here, $n_{i\alpha} =
c_{i\sigma}^{\dagger}c_{i\sigma}$ and $c_{i\sigma}^\dagger$
($c_{i,\sigma}$) creates (destroys) an electron of spin
projection $\sigma$ at site $i$. The sum $<i,j>$ is over nearest neighbors and
the hoping $t$ is taken to be a constant. A band structure analysis
shows that the conduction band touches the valence band at discrete
points, so the low-energy excitations are massless Dirac fermions.
Around the touching Dirac points, the kinetic term can be converted
to a free Lagrangian term
\begin{equation}
\mathcal{L}_0 = \sum_{\sigma}
\bar{\Psi}_{\sigma}(\partial_{\tau}\gamma_0 - i
v_{F}\partial_{x}\gamma_1 - i
v_{F}\partial_{y}\gamma_2)\Psi_{\sigma},
\end{equation}
where $\gamma_{0}=\mathbbm{1}\otimes\tau_{3}$,
$\gamma_{1}=\mathbbm{1}\otimes i\tau_{1}$, and
$\gamma_{2}=\tau_{3}\otimes i\tau_{2}$. The $4$-component spinor
$\Psi_{\sigma}$ represents Dirac fermion, and is arranged as
$\Psi_{\sigma}=(c_{AK\sigma}, c_{BK\sigma}, c_{AK'\sigma},
c_{BK'\sigma})^{T}$ with sublattice indices $A, B$ and valley
indices $K, K'$. Its conjugate is $\bar{\Psi}_{\sigma} =
\Psi^{\dagger}_{\sigma}\gamma_0$. The on-site Hubbard interaction
can be rewritten as the sum of a number of continuum four-fermion
interaction terms, among which only the following term turns out to
be relevant in the RG sense \cite{Herbut06},
\begin{equation}
\mathcal{L}_{\mathrm{int}} = -\frac{2g}{N}\Big(\sum_{\sigma}
\sigma\bar{\Psi}_{\sigma}(\tau,\mathbf{r})
\Psi_{\sigma}(\tau,\mathbf{r})\Big)^2,
\end{equation}
where $g = Ua^2/8$ and $a$ is the lattice constant.

We first sketch Hertz's approach~\cite{Hertz}. It is convenient to
introduce a scalar field $\phi$ and perform a Hubbard-Stratonovich
transformation, yielding
\begin{eqnarray}
\mathcal{L} = \mathcal{L}_0 +
\frac{N}{2}\left[(\partial_{\tau}\phi)^{2} +
(\mathbf{\nabla}\phi)^{2} +r\phi^{2}\right] -
\phi\sum_{\sigma}\sigma \bar{\Psi}_{\sigma}\Psi_{\sigma}.
\end{eqnarray}
Integrating out the fermion fields $\bar{\Psi},\Psi$ and taking the
saddle point approximation leads to a gap equation \cite{Rosenstein}
\begin{equation}
\frac{\langle\phi \rangle}{g} = 16\int_{\omega,\mathbf{k}}
\frac{\langle\phi \rangle}{\omega^2+{\mathbf{k}}^2 + \langle \phi
\rangle^2}.
\end{equation}
The critical point separating zero and finite expectation value
$\langle\phi \rangle$ is given by $\frac{1}{g_c} = 16
\int_{\omega,\mathbf{k}} \frac{1}{\omega^2+\mathbf{k}^{2}}$. For $g
< g_{c}$, the system is in the gapless SM phase with
$\langle\phi\rangle = 0$. For $g > g_{c}$, $\langle\phi\rangle \neq
0$, so the Dirac fermions acquire a finite mass gap and the system
is turned to a Mott insulator. Such excitonic-type insulating
transition is a condensed-matter analog of the dynamical
fermion-mass generation mechanism proposed originally in particle
physics \cite{Nambu}. The order parameter can be expressed in terms
of fermion operators as $\langle \phi \rangle = \langle\sum_{\sigma}
\sigma \bar{\Psi}_{\sigma} \Psi_{\sigma}\rangle$, so the Mott
insulator is indeed an AFM. Close to $g_c$, the fermion gap is
approximated by $\langle \phi \rangle = \frac{\pi}{4}
\left(\frac{1}{g_c} - \frac{1}{g}\right)$, which suggests that the
SM-AFM transition is continuous. This phase transition is tuned by
the parameter $r = \frac{1}{4}\left(\frac{1}{g} -
\frac{1}{g_c}\right)$, with $g_c$ defining the zero-temperature
quantum critical point.

\begin{figure}[t!]
   \includegraphics[width=3in]{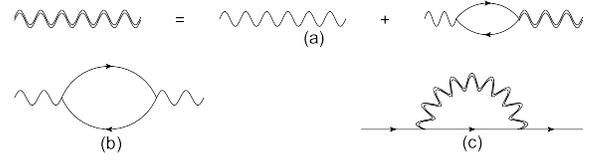}
\caption{(a) the dressed boson propagator; (b) the one-loop diagram
for the polarization function; (c) the fermion self-energy
diagram.}\label{self}
\end{figure}

We will now demonstrate that Hertz' scenario of the SM-AFM
transition on the honeycomb lattice can be fundamentally altered by
the coupling between Dirac fermions and critical order parameter
fluctuations. This coupling describes the decay of composite order
parameter into single fermionic excitations as well as the formation
of order parameter by fermion pairing. Since it affects the dynamics
of both fermions and bosons, we are led to maintain the fermions in
the Yukawa-type theory, Eq.~(4), and study the fermions and bosons
on equal footings. This will be done by analyzing the Yukawa
interaction by means of an $1/N$ expansion where $N$ represents the
fermion flavor (spin).

Due to the Lorentz invariance of the action in the low-energy
sector, the dynamical exponent is $z=1$. The free fermion propagator
is $G_0(\omega,\mathbf{k}) =
\frac{1}{i\omega\gamma_0-\mathbf{\gamma\cdot k}}$. In leading order
of $1/N$ expansion, the polarization is given by
\begin{eqnarray}
\Pi_0(\Omega,\mathbf{q}) = \int_{\omega,\mathbf{k}}
\mathrm{Tr}\left[G_0(\omega,\mathbf{k})
G_0(\omega+\Omega,\mathbf{k+q})\right],
\end{eqnarray}
and shown in Fig.~1(b). It follows that $\Pi_0(\Omega,\mathbf{q}) =
\frac{1}{4}\sqrt{\Omega^2 + \mathbf{q}^2}$. The bosonic propagator
at the same order in $1/N$ is therefore $D^{-1}(\Omega,\mathbf{q}) =
N\left[\Omega^2 + q^2 + \Pi_0(\Omega,\mathbf{q})\right]$. In the low
energy regime, the free terms turn out to be unimportant, so
$D(\Omega,\mathbf{q}) = \frac{4}{N\sqrt{\Omega^2 + \mathbf{q}^2}}$.
The leading fermion self-energy diagram is shown in Fig. 1(c).
Retaining the most divergent part leads to
$\Sigma_0(\omega,\mathbf{k}) = - \frac{2}{3\pi^2 N}(i\omega\gamma_0
- \mathbf{\gamma\cdot k}) \ln\left(\frac{\Lambda}{\sqrt{\omega^2 +
\mathbf{k}^2}}\right)$. From the Dyson equation,
$G^{-1}(\omega,\mathbf{k}) = i\omega\gamma_0 - \mathbf{\gamma\cdot
k} - \Sigma_0(\omega,\mathbf{k})$, this results in  a dressed
fermion propagator
\begin{eqnarray}
G(\omega,\mathbf{k}) = -\frac{i\omega\gamma_0 - \mathbf{\gamma \cdot
k}}{(\omega^2 + \mathbf{k}^2)^{1-\frac{\eta}{2}}},
\label{eqn:DressedFermion}
\end{eqnarray}
with anomalous dimension $\eta = \frac{2}{3\pi^2N}$. This propagator
implies the absence of a well-defined quasiparticle pole.

So far, our calculation is on the level of random phase
approximation since the polarization $\Pi_0(\Omega,\mathbf{q})$ has
been obtained from the free propagator $G_0(\omega,\mathbf{k})$.
However, $\Pi_0(\Omega,\mathbf{q})$ may be considerably modified
when the feedback of fermion dressing is included. Replacing
$G_0(\omega,\mathbf{k})$ by $G(\omega,\mathbf{k})$ shown in
Eq.(\ref{eqn:DressedFermion}), we obtain a new polarization
\begin{eqnarray}
\Pi_0^{\prime}(\Omega,\mathbf{q}) =  A(\eta)(\Omega^2 +
\mathbf{q}^2)^{\frac{1}{2}+\eta}, \label{eqn:DressedPola}
\end{eqnarray}
where $A(\eta)$ is a well-defined function of
$\eta$~\footnote{$A(\eta) =
\frac{\Gamma(\frac{1}{2}-\eta)-\frac{3}{2}\Gamma(-\frac{1}{2}-\eta)}
{2\pi^{\frac{3}{2}}\Gamma(1-\frac{\eta}{2})\Gamma(1-\frac{\eta}{2})}
\int^1_0 dx \left[x(1-x)\right]^{\frac{1}{2}+\frac{\eta}{2}}.$}. The
fermion self-energy and anomalous dimension $\eta$ may also be
altered, so it is necessary to recalculate the fermion self-energy
by including the dressing effects of both fermions and bosons. Using
the dressed expressions, Eqs.~(7) and (8), we get a new fermion
self-energy $\Sigma(\omega,\mathbf{k}) = -\frac{B(\eta)}{8\pi^3
NA(\eta)}\frac{i\omega\gamma_0 - \mathbf{\gamma\cdot k}}{(\omega^2 +
\mathbf{k}^2)^{\frac{\eta}{2}}}$, where $B(\eta)$ is another
function of $\eta$~\footnote{$ B(\eta) = \frac{4\pi
\Gamma\left(\frac{3}{2}+\frac{\eta}{2}\right)}
{\Gamma\left(\frac{1}{2}+\eta\right)\Gamma\left(1-\frac{\eta}{2}\right)}\int_0^1dx
\int_{0}^{+\infty}dy
\frac{x^{\frac{1}{2}+\eta}(1-x)^{-\frac{\eta}{2}} y^2}
{\left[y^2+x\left(1-x\right)k^2\right]^{\frac{3}{2}+\frac{\eta}{2}}}.$}.
Replacing $\Sigma_0(\omega,\mathbf{k})$ by
$\Sigma(\omega,\mathbf{k})$, we are again lead to
Eq.(\ref{eqn:DressedFermion}). The dressed fermion propagator
$G(\omega,\mathbf{k})$ thus survives the self-consistency analysis.

We are now in a position to examine the validity of Hertz's $\phi^4$
theory and the stability of the continuous SM-AFM quantum phase
transition. According to Hertz \cite{Hertz}, one can formally
integrate out all fermionic degrees of freedom and obtain an
effective action as a function of the order parameter $\phi$:
\begin{eqnarray}
S_{\mathrm{eff}}[\phi] = \int_{\Omega,\mathbf{q}}
\left[\chi_2(\Omega,\mathbf{q})\phi^{2} + \chi_4(\Omega,\mathbf{q})
\phi^4 + O(\phi^{6})\right].
\end{eqnarray}
Here, the quadratic coefficient is determined by the spin
susceptibility: $\chi_2(\Omega,\mathbf{q}) = \Omega^2 + \mathbf{q}^2
+ \Pi(\Omega,\mathbf{q})$. The validity of this $\phi^4$ theory is
rooted on the basic assumption that the coefficients
$\chi_2(\Omega,\mathbf{q})$ and $\chi_4(\Omega,\mathbf{q})$ are
regular in $\Omega, \mathbf{q}$.

We first consider the coefficient of the quartic $\phi^4$ term.
According to the leading diagram shown in Fig. 2, it is
\begin{eqnarray}
\chi_4(\Omega,\mathbf{q}) &=& \int \frac{d\omega d^2
\mathbf{k}}{(2\pi)^3}\mathrm{Tr}
\left[G(\omega,\mathbf{k})G(\omega+\Omega,\mathbf{k+q})\right.
\nonumber \\
&& \left.\times
G(\omega,\mathbf{k})G(\omega+\Omega,\mathbf{k+q})\right].
\end{eqnarray}
Using the dressed fermion propagator Eq.(\ref{eqn:DressedFermion})
and carrying out some tedious calculations, one can show that
\begin{eqnarray}
\chi_4(\Omega,\mathbf{q}) = \frac{F(\eta)}{(\Omega^2 +
\mathbf{q}^2)^{\frac{1}{2} - 2\eta}},\label{eqn:PrefactorPhi4}
\end{eqnarray}
with
\begin{eqnarray}
F(\eta) &\propto& \frac{2\Gamma(4 -
2\eta)}{\Gamma^2(2-\eta)}\left[\frac{3}{2}
\frac{\Gamma(\frac{1}{2}-2\eta)}{\Gamma(3-2\eta)} +
\frac{\Gamma(\frac{5}{2}-2\eta)}{\Gamma(4-2\eta)} \right. \nonumber \\
&& \left.-\frac{13}{2}
\frac{\Gamma(\frac{3}{2}-2\eta)}{\Gamma(4-2\eta)}\right]
\int_{0}^{1}\frac{dx}{[x(1-x)]^{-\frac{1}{2}-\eta}} \nonumber \\
&& + \left[\frac{\Gamma(\frac{3}{2}-2\eta)} {\Gamma^2(2-\eta)} -
\frac{\Gamma(\frac{1}{2}-2\eta)}{\Gamma^2(1-\eta)}\right]
\int_{0}^{1} \frac{dx}{[x(1-x)]^{\frac{1}{2}-\eta}}. \nonumber
\end{eqnarray}
The function $F(\eta)$ is finite for any finite $\eta$ and it is
easy to see that $\chi_4(\Omega,\mathbf{q})$ is divergent in the
$\Omega,\mathbf{q} \rightarrow 0$ limit. In other words, the quartic
$\phi^4$ term is singular. An important feature of $F(\eta)$ is that
it vanishes in the absence of fermion anomalous dimension, i.e.,
$F(\eta = 0) = 0$, thus, this singularity is absent if the free
propagator $G_0(\omega,\mathbf{k})$ is used. Therefore, the
singularity in $\chi_4(\Omega,\mathbf{q})$ arises directly from the
critical order parameter fluctuations, which is reflected in the
finiteness of $\eta$.

\begin{figure}[t!]
    \includegraphics[width=0.8in]{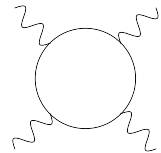}
 \caption{$\phi^4$ vertex correction at the one-loop level.}
     \label{fig:Ix}
\end{figure}

\begin{figure}[t!]
\vspace{-1.5cm} \hspace{-0.34cm}
   \includegraphics[width=3.3in]{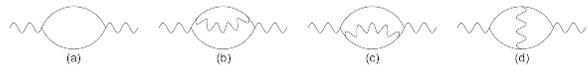}
\caption{Polarization functions up to the two-loop level.}
\end{figure}

We now turn to analyze the spin susceptibility at the quantum
critical point. For the effective $\phi^4$ theory to be applicable
and the continuous SM-AFM transition to be stable, the static spin
susceptibility $\chi_2(\mathbf{q})$ should have a regular momentum
dependence. It is commonly believed that the spin susceptibility of
a Fermi liquid is regular. Interestingly, Belitz \emph{et}
\emph{al.} demonstrated that a nonanalytic spin susceptibility in a clean
Fermi liquid system can occur~\cite{Belitz}. Subsequently, Chubukov \emph{et}
\emph{al.} obtained a negative, nonanalytic spin susceptibility for a
ferromagnetic quantum phase transition system with dynamical
exponent $z=3$ \cite{Chubukov}. It is therefore necessary to
check whether the spin susceptibility is regular or singular.

The spin susceptibility actually corresponds to the polarization
function $\Pi(q)$. The leading diagram is shown as Fig. 3(a), with
$\Pi_{2}^{a}(\Omega,\mathbf{q})$ being already given by Eq.(8). The
two-loop contributions are represented by the three diagrams (b),
(c), and (d) of Fig. 3. Using Eqs.~(7) and (8), we find that the sum of
(b) and (c) is
\begin{eqnarray}
\Pi_{2}^{b+c}(\Omega,\mathbf{q}) = -2C_{1}(\eta)
\Pi_{2}^{a}(\Omega,\mathbf{q}),
\end{eqnarray}
where $C_1(\eta) = \frac{B(\eta)}{8\pi^3NA(\eta)}$. Employing a
method used in \cite{Franz} and then performing a lengthy
computation, we find
\begin{eqnarray}
\Pi_{2}^{d}(\Omega,\mathbf{q}) = -2C_{2}(\eta)
\ln\Big(\frac{\Lambda}{\sqrt{\Omega^2 + \mathbf{q}^2}}\Big)
\Pi_{2}^{a}(\Omega,\mathbf{q}),
\end{eqnarray}
where $C_{2}(\eta)$ is a function of $\eta$. Adding
$\Pi_{2}^{a}$, $\Pi_{2}^{b+c}$, and $\Pi_{2}^{d}$ together one arrives at the expression
\begin{eqnarray}
\Pi_2(\Omega,\mathbf{q}) &=&
\left(\Omega^2+\mathbf{q}^2\right)^{-\frac{\eta'}{2}}
\Pi_{2}^{a}(\Omega,\mathbf{q}) -
2C_{1}(\eta)\Pi_{2}^{a}(\Omega,\mathbf{q}) \nonumber \\
&\propto& \Big[\left(\Omega^2+\mathbf{q}^2
\right)^{-\frac{\eta'}{2}} - 2 C_{1}(\eta)\Big]
\left(\Omega^2+\mathbf{q}^2\right)^{\frac{1}{2}+\eta},
\nonumber \\
\end{eqnarray}
where $\eta' = -2C_{2}(\eta)$. Since $\eta' < 0$, the function
$\Pi_2(\Omega,\mathbf{q})$ is positive for large $\Omega,\mathbf{q}$
but becomes negative for $\Omega,\mathbf{q}$ below a given
threshold. As a result, the corresponding spin susceptibility
$\chi_2(\Omega,\mathbf{q})$ is non-analytical.

As with the $\chi_4$ term, this non-analyticity is directly related
to the fermion anomalous dimension: If we assume $\eta=0$ and apply
the free fermion propagator $G_0(\omega,\mathbf{k})$, then
$\Pi_{2}^a(\Omega,\mathbf{q})$ is identical to
$\Pi^{0}(\Omega,\mathbf{q})$ given by Eq.(6), and both
$\Pi_{2}^{b+c}(\Omega,\mathbf{q})$ and
$\Pi_{2}^{d}(\Omega,\mathbf{q})$ are proportional to
$\ln\Big(\frac{\Lambda}{\sqrt{\Omega^2 + \mathbf{q}^2}}\Big)
\Pi^0(\Omega,\mathbf{q})$. The total polarization for $\eta = 0$
will be $\Pi_2(\Omega,\mathbf{q},\eta = 0) \propto
(\Omega^2+\mathbf{q}^2)^{\frac{1}{2} + \frac{8}{3\pi^2 N}}$, which
is positive and analytical.

Based on the above analysis, we arrive at a key conclusion: the
non-analyticities in $\chi_4(\Omega,\mathbf{q})$ and
$\chi_2(\Omega,\mathbf{q})$ clearly signal the breakdown of Hertz's
$\phi^4$ theory. Moreover, the non-analytical spin susceptibility
will alter the nature of the SM-AFM transition. To demonstrate this,
we write the effective action for $\phi$ as $S_{\mathrm{eff}}[\phi]
= \int_{q} \left[\left(q^2 + \Pi_2(q)\right)\phi^{2} + \chi_4(q)
\phi^4 \right]$ with $\Pi_2(q)$ given by Eq.(14). To determine its
ground state, one needs to find the minimum of the total energy. The
absolute minimum of kinetic energy $E_{q} = q^2 + \Pi_2(q)$ is
located at a finite momentum $q_c$, which is one of the solutions of
equation $d E_{q}/dq = 0$. Though it is hard to derive an analytical
expression of $q_c$, its finiteness can be easily confirmed by
numerical computation. Inserting $q_c$ to $E_q$, we find that the
non-analytical spin susceptibility leads to a negative minimal
kinetic energy, i.e., $E_{q_c}^{\mathrm{min}} < 0$. Around $q_c$,
$E_{q}$ can be approximated by $E_{q} \approx E_{q_c}^{\mathrm{min}}
+ \kappa(q-q_c)^2$ with certain constant $\kappa$. Originally, at
the critical point $g = g_c$, the scalar field $\phi$ has a
vanishing mass, $r = 0$, so the potential energy is simply $V[\phi]
\propto \phi^4$ and one should expect $\langle \phi \rangle = 0$.
However, the finite $E_{q_c}^{\mathrm{min}}$ induced by
non-analytical spin susceptibility serves as an effective, negative
mass. Moreover, $\chi_4(q)$ is regular for finite $q_c$, and hence
can be safely replaced by a constant. Consequently, the effective
potential becomes $V[\phi] = -\left|E_{q_c}^{\mathrm{min}}\right|
\phi^{2} + \chi_4(q_c) \phi^4$, which yields a finite expectation
value, $\langle \phi \rangle =
\left[-E_{q_c}^{\mathrm{min}}/2\chi_4(q_c)\right]^{\frac{1}{2}}$,
even if $r=0$. According to the generic analysis of Brazovskii
\cite{Brazovskii}, this nonzero order parameter $\langle \phi
\rangle$ should be nonuniform. By minimizing the corresponding free
energy, we find the order parameter is spatially modulated as
$\phi(\mathbf{x}) = \phi_0 e^{i\mathbf{q}_c \cdot \mathbf{x}}$.

In the SM phase with $r>0$, the order parameter fluctuation is
gapped, so the Yukawa interaction does not generate any fermion
anomalous dimension. As explained in the above discussions,
$\chi_2(\omega,\mathbf{q})$ and $\chi_4(\omega,\mathbf{q})$ exhibit
no singular behaviors when $\eta = 0$, thus the spin susceptibility
is analytical for $r > 0$ and non-analytical only at the critical
point $r=0$. Therefore, $\langle \phi \rangle$ vanishes for any
positive $r$, but develops a finite magnitude at $r = 0$
discontinuously because of the negativeness of
$E_{q_c}^{\mathrm{min}}$. Apparently, although SM-AFM phase
transition is shown to be continuous by the gap equation analysis,
it is ultimately driven first order due to singular interaction
between order parameter and massless Dirac fermions.

In summary, we analyzed the nature of the quantum phase transition
between a semimetal and an antiferromagnet on a honeycomb lattice.
We demonstrated that the strong interaction between massless Dirac
fermions and critical order parameter fluctuations generates
unexpected properties. As a consequence, the original Hertz $\phi^4$
theory is incomplete for this problem and the massless fermions
should be treated on equal footing with the critically fluctuating
order parameter. This behavior arises from the non-vanishing fermion
anomalous dimension, which in turn reflects strong damping effects
that need to be included. Our results indicate that, although the
SM-AFM transition is continuous at the mean-field level, it is
destroyed by the critical order parameter fluctuations and actually
gives way to a first order transition with a spatially modulated
order parameter. The anticipated quantum critical phenomena of
continuous SM-AFM transition \cite{Sachdev, Herbut06, Herbut09,
Fritz} are not expected to exist.

G.Z.L. acknowledges support by the National Natural Science
Foundation of China under grant No.11074234 and the Visitors Program
of MPIPKS at Dresden.


\begin{thebibliography}{99}

\bibitem{understand}
\emph{Understanding Quantum Phase Transitions}, edited by L. D. Carr
(CRC Press, 2011).

\bibitem{Sachdevbook}
S. Sachdev, \emph{Quantum Phase Transitions} (Cambridge University
Press, 2000).

\bibitem{Vojta}
Ar. Abanov \emph{et} \emph{al.}, Adv. Phys. {\bf 52}, 119 (2003); M.
Vojta, Rep. Prog. Phys. {\bf 66}, 2069 (2003); D. Belitz \emph{et}
\emph{al.}, Rev. Mod. Phys. {\bf 77}, 579 (2005); H. v.
L$\ddot{\mathrm{o}}$hneysen \emph{et} \emph{al.}, Rev. Mod. Phys.
{\bf 79}, 1015 (2008); P. Gegenwart \emph{et} \emph{al.}, Nat. Phys.
{\bf 4}, 186 (2008).

\bibitem{Sachdev}
S. Sachdev, arXiv:1012.0299v3.

\bibitem{Herbut06}
I. F. Herbut, Phys. Rev. Lett. {\bf 97}, 146401 (2006).

\bibitem{Herbut09}
I. F. Herbut \emph{et} \emph{al.}, Phys. Rev. B {\bf 79}, 085116
(2009); Phys. Rev. B {\bf 80}, 075432 (2009).

\bibitem{Fritz}
L. Fritz, Phys. Rev. B {\bf 83}, 035125 (2011).

\bibitem{Honerkamp}
C. Honerkamp, Phys. Rev. Lett. {\bf 100}, 146404 (2008).

\bibitem{Sorella}
S. Sorella and E. Tosatti, Europhys. Lett. {\bf 19}, 699 (1992); T.
Pavia \emph{et} \emph{al.}, Phys. Rev. B {\bf 72}, 085123 (2005).

\bibitem{Hermele}
M. Hermele, Phys. Rev. B {\bf 76}, 035125 (2007).

\bibitem{Wangfa}
F. Wang, Phys. Rev. B {\bf 82}, 024419 (2010).

\bibitem{Ran}
Y. M. Lu and Y. Ran, Phys. Rev. B {\bf 84}, 024420 (2011).

\bibitem{Wen}
A. Vaezi and X.-G. Wen, arXiv:1010.5744.

\bibitem{Meng}
Z. Y. Meng \emph{et} \emph{al.}, Nature (London) {\bf 464}, 847
(2010).

\bibitem{Hertz}
J. Hertz, Phys. Rev. B {\bf 14}, 1165 (1976).

\bibitem{Millis}
A. J. Millis, Phys. Rev. B {\bf 48}, 7183 (1993).

\bibitem{Moriya}
T. Moriya, \emph{Spin Fluctuations in Itinerant Electron Magnetism}
(Springer-Verlag, Berlin, New York, 1995).

\bibitem{Belitz}
D. Belitz \emph{et} \emph{al.}, Phys. Rev. B {\bf 55}, 9452 (1997).

\bibitem{Chubukov}
A. V. Chubukov \emph{et} \emph{al.}, Phys. Rev. Lett. {\bf 92},
147003 (2004); J. Rech \emph{et} \emph{al.}, Phys. Rev. B {\bf 74},
195126 (2006).

\bibitem{Abanov}
A. Abanov and A. V. Chubukov, Phys. Rev. Lett. {\bf 93}, 255702
(2004).

\bibitem{Rosenstein}
B. Rosenstein \emph{et al.}, Phys. Rep. {\bf 205}, 59 (1991).

\bibitem{Nambu}
Y. Nambu and G. Jona-Lasinio, Phys. Rev. {\bf 122}, 345 (1961).

\bibitem{Franz}
M. Franz \emph{et} \emph{al.}, Phys. Rev. B {\bf 68}, 024518 (2003).

\bibitem{Brazovskii}
S. A. Brazovskii, Sov. Phys. JETP {\bf 41}, 85 (1975).


\end{thebibliography}
\end{document}